\documentclass[aps,prl,twocolumn,groupedaddress]{revtex4-1}

\usepackage{amsmath}
\usepackage{amssymb}
\usepackage{hyperref}
\usepackage{graphicx}
\usepackage{standalone}
\usepackage{subfig}
\usepackage{enumitem}

\begin{document}


\title{Graph's Topology and Free Energy of a Spin Model on the Graph}


\author{Jeong-Mo Choi}
\email[]{jeongmochoi@fas.harvard.edu}
\author{Amy I. Gilson}
\email[]{amygilson@fas.harvard.edu}
\author{Eugene I. Shakhnovich}
\email[]{shakhnovich@chemistry.harvard.edu}
\affiliation{Department of Chemistry and Chemical Biology, Harvard University, 12 Oxford Street, Cambridge, Massachusetts 02138, USA}


\date{\today}

\begin{abstract}
In this work we show that there is a direct relationship between a graph's topology and the free energy of a spin system on the graph. We develop a method of separating topological and enthalpic contributions to the free energy, and find that considering the topology is sufficient to qualitatively compare the free energies of different graph systems at high temperature, even when the energetics are not fully known. This method was applied to the metal lattice system with defects, and we found that it partially explains why point defects are more stable than high-dimensional defects. Given the energetics, we can even quantitatively compare free energies of different graph structures via a closed form of linear graph contributions. The closed form is applied to predict the sequence space free energy of lattice proteins, which is a key factor determining the designability of a protein structure.
\end{abstract}


\maketitle


Over the last thirty years, graph theory has been applied to the study of various networks, including protein interaction networks, neural networks, and the World Wide Web \cite{Watts:1998rz, RevModPhys.74.47, pastor2007evolution, dorogovtsev2013evolution}. Especially the interplay between network structure and dynamics has attracted considerable attention \cite{RevModPhys.80.1275}, while the equilibrium characteristics of networks, which may deepen our understanding of network phenomena, have not yet been studied thoroughly \cite{Farkas2004}.

In this work, we will consider a spin model on a graph, which has a wide range of applications from gene expression on biochemical networks \cite{Konig01072004} to social network phenomena \cite{Bisconti:2015fk}, and study an analytical relationship between a system's graph topology and its free energy. Such a relationship would be useful when discriminating between topologies based on their stability, and it could even provide the insight into graph dynamics, although we only consider systems with fixed graph topology here.

Consider a simple graph \footnote{A simple graph is defined as a graph with no self-loops on any node and no mulitple links between any pair of nodes. It may be either connected or disconnected.} of $N$ nodes. The graph connectivity is described by the adjacency matrix $A$, whose element $A_{ij}$ is 1 when there is a link between nodes $i$ and $j$, and $A_{ij} = 0$ otherwise. Each node is in one of $M$ possible spin states. The Hamiltonian, $\mathcal{H}$, is defined as the summation of energetic contributions over all links, each of whose energy is determined by the states of its two terminal nodes. Note that orphan nodes do not contribute energetically, by definition. Now, the Hamiltonian can be written formally as
\begin{eqnarray} \label{eq:hamil}
\mathcal{H} &=& \frac{1}{2} \sum_{i,j}^{N,N} A_{ij} E_{s(i) s(j)},
\end{eqnarray}
where $E$ is the energy matrix and $s(i)$ is the state of node $i$. The high-temperature expansion of the partition function $Z(\beta) = \sum_{\{s\}} e^{-\beta \mathcal{H}}$ over all possible state configurations is
\begin{equation}
Z(\beta) = \sum_{\{s\}} 1 - \beta \sum_{\{s\}} \mathcal{H} + \frac{\beta^2}{2!} \sum_{\{s\}} \mathcal{H}^2 - \cdots.
\end{equation}

\newcommand{\graphone}{\ensuremath{\vcenter{\hbox{\includegraphics[height=5ex]{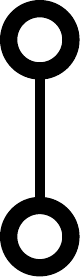}}}}}

\newcommand{\graphtwoa}{\ensuremath{\vcenter{\hbox{\includegraphics[height=5ex]{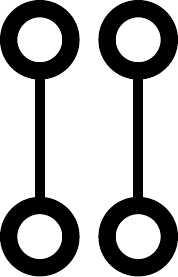}}}}}
\newcommand{\graphtwob}{\ensuremath{\vcenter{\hbox{\includegraphics[height=5ex]{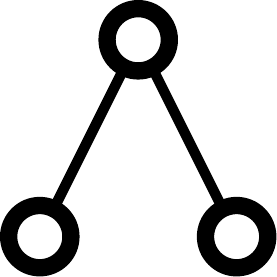}}}}}
\newcommand{\graphtwoc}{\ensuremath{\vcenter{\hbox{\includegraphics[height=5ex]{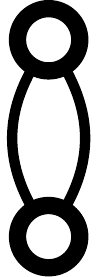}}}}}

\newcommand{\graphthreef}{\ensuremath{\vcenter{\hbox{\includegraphics[height=5ex]{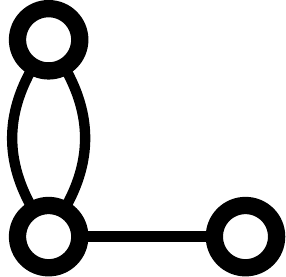}}}}}

\begin{figure}[tb]
\subfloat[] {
\includegraphics[scale=0.5]{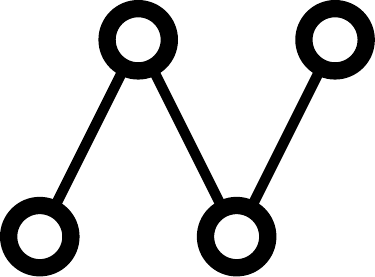}
} \qquad
\subfloat[] {
\includegraphics[scale=0.5]{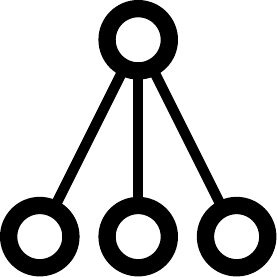}
} \qquad
\subfloat[] {
\includegraphics[scale=0.5]{Fig_Obeta3f}
} \qquad
\caption{\label{fig:obeta3} Examples of 3-link multigraphs. a is a parent graph of c (\emph{i.e.}, c is obtained by a node contraction operation on a), and b is also a parent graph of c, but there is no such relationship between a and b.}
\end{figure}

In the higher-order terms, there are summations of various products of $A$ and $E$ elements. To systematically visualize those terms, we introduce a multigraph $g$ (where multiple links between any pair of nodes are allowed) with no orphan nodes (see Fig. \ref{fig:obeta3} for examples), and we use its nodes as dummy variables of the summations. For each multigraph $g$, we define two quantities $[g]$ and $E(g)$ as follows:
\begin{eqnarray}
[g] &=& \sum_\text{nodes} \prod_{k=1}^{n(g)} A_{l_k}, \\
E(g) &=& M^{-n(\text{nodes})} \sum_\text{nodes} \prod_{k=1}^{n(g)} E_{l_k},
\end{eqnarray}
where $l_k$ indicates the link of index $k$, $n(g)$ is the number of links in graph $g$, and $n(\text{nodes})$ is the number of nodes in graph $g$. For example, for the graph shown in Fig. \ref{fig:obeta3}c,
\begin{eqnarray} \label{eq:ex1}
\left[\graphthreef\right] &=& \sum_{ijk} A_{ij} A_{ij} A_{jk}, \\
E\left(\graphthreef\right) &=& M^{-3} \sum_{mnp} E_{mn} E_{mn} E_{np}.
\end{eqnarray}
Note that since $A$ is a binary matrix, $[g]$ for a graph $g$ with multiple links is equal to $[g_0]$ where $g_0$ is a simple graph constructed from $g$ by converting all multiple links in $g$ to single links. For example, equation \ref{eq:ex1} becomes
\begin{equation}
\left[\graphthreef\right] = \sum_{ijk} A_{ij}^2 A_{jk} = \sum_{ijk} A_{ij} A_{jk} = \left[\graphtwob\right].
\end{equation}

Next, we introduce a special form of graph operation: node contraction, which merges two different nodes while preserving the number of links \cite{hartung2015}. If a graph $h$ can be obtained by any number of node contraction operations on another graph $g$, we will call $h$ a child graph of $g$, and $g$ a parent graph of $h$. For example, in Fig. \ref{fig:obeta3}, graph a is a parent graph of c, and graph b is also a parent graph of c, but there is no parent-child relationship between a and b. Note that $[g]$ represents the total number of unique subgraphs of type $g$ and of its child types on the given graph.

Then it can be shown (see \hyperref[subsec:appA]{\textbf{Appendix A}}) that
\begin{equation} \label{eq:Zorig}
Z(\beta) = M^N \exp \left\{ \sum_{\text{connected } g} \frac{(-\beta/2)^{n(g)}}{n(g)!} H(g)[g]\right\},
\end{equation}
where the summation is over all possible connected subgraphs $g$. $H(g)$ is defined as
\begin{eqnarray} \nonumber
H(g) &=& K(g) E(g) \\ \label{eq:Hg}
&& +\sum_{g' \in \mathcal{P}(g)} (-1)^{m(g,g')} K(g,g') K(g') E(g'),
\end{eqnarray}
where $K(g)$ is the combinatoric factor to construct graph $g$ from $n(g)$ links, $K(g, g')$ is the combinatoric factor to generate graph $g$ from graph $g'$ by node contraction, $m(g, g')$ is the number of contraction operations required to construct $g$ from $g'$, and $\mathcal{P}(g)$ is the set containing all parent graphs of graph $g$. Finally, the free energy is
\begin{equation} \label{eq:FE}
F(\beta) = - N k_B T \ln M + \sum_{\text{connected } g} \tilde{F}(g, \beta),
\end{equation}
where
\begin{equation} \label{eq:Fgb}
\tilde{F}(g,\beta) = -\frac{1}{\beta} \frac{(-\beta/2)^{n(g)}}{n(g)!} H(g) [g].
\end{equation}

%

One advantage of equations \ref{eq:FE} and \ref{eq:Fgb} is that the graph topology (which determines $[g]$) is now unlinked from detailed energetics (which determines $H(g)$). Hence, even without knowing the exact energy matrix $E$, it is possible to compare $[g]$ values from different structures and, in some cases, we can determine which structure provides lower free energy of the corresponding spin system. To illustrate this, let us consider two different graph systems, a chain graph of length $N$ and a star graph with $N$ leaves. They have the same numbers of nodes and links, but it can be shown that $[g]^\text{chain} < [g]^\text{star}$ holds in general (see \hyperref[subsec:appB]{\textbf{Appendix B}}), so at a temperature high enough that, for the star graph, the infinite sum in equation \ref{eq:FE} does not diverge and if the sum is negative (stable free energy), we can conclude that the star-graph spin system has lower free energy than its counterpart on a chain graph. Note that this qualitative result is independent of the details of the energy matrix. As a specific example, an Ising chain system always has higher free energy than an Ising star at any $T > 0$ regardless of the details of energetics (see \hyperref[subsec:appB]{\textbf{Appendix B}}). 


\begin{figure}[tb]
\includegraphics[width=0.5\textwidth]{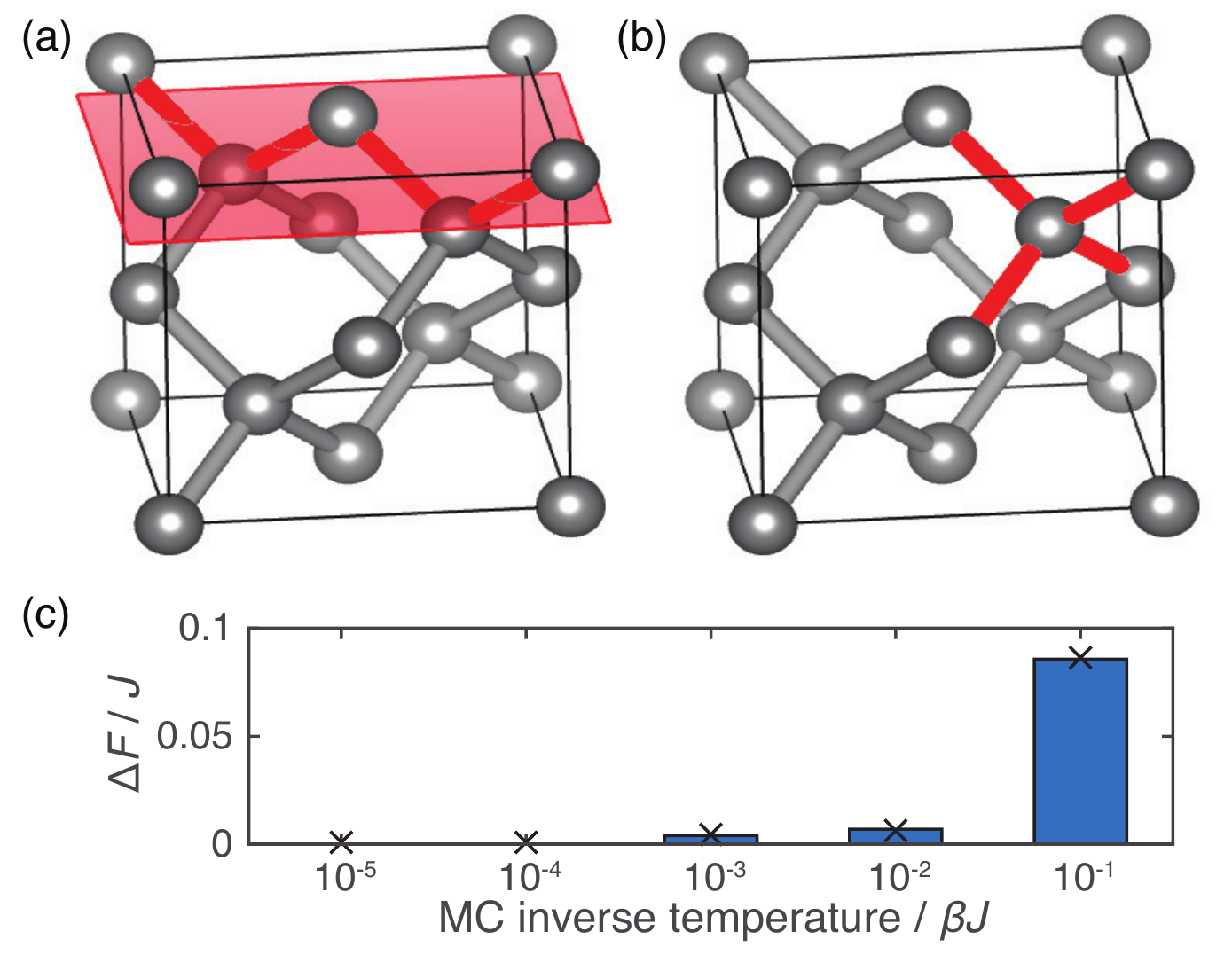}
\caption{\label{fig:alloystr} (a-b) Two different types of defects, drawn by VESTA 3 \cite{Momma:db5098}. Broken bonds are marked in red. (a) Diamond structure with a grain boundary indicated by a red plane. (b) Same structure with a vacancy defect. (c) Difference between MC free energies of two Si-Ge alloy lattice systems with different types of defects, as a function of inverse temperature. Here, the $y$ axis shows $\Delta F = F(\text{vacancy defect}) - F(\text{grain boundary})$ for the entire system.}
\end{figure}

A possible application of this general analysis is to an alloy lattice system with defects. Here we consider two types of defects: planar defects (\emph{i.e.}, grain boundaries) and point defects. We constructed a 3-dimensional diamond-like lattice structure in 3 $\times$ 3 $\times$ 3 unit cells with a periodic boundary condition (216 lattice points). The first system contains a grain boundary modeled by a discontinuity on the (001) plane (see Fig. \ref{fig:alloystr}a). To simulate vacancy defects, we constructed a second system by removing lattice points randomly (see Fig. \ref{fig:alloystr}b) until the number of broken bonds was equal to the number of bonds broken at the grain boundary in the first system (the number of remaining bonds = 324). Using a similar argument as above, it can be analytically shown that $[g]$ is generally greater for the point vacancy system than for the grain boundary system.

We carried out a Monte Carlo (MC) simulation to check that the system with point vacancies indeed has lower free energy than that with a grain boundary. We considered lattice site occupation with the atom types $s(i) = \text{Si}, \text{Ge}$ as the possible states of lattice site $i$, and used the interatomic potential developed in previous works \cite{PhysRevB.51.4894, PhysRevB.71.134104}. Assuming equal bond lengths, $E(\text{Si},\text{Si}) = -2.17 J, E(\text{Ge},\text{Ge}) = -1.93 J, E(\text{Si},\text{Ge}) = -2.04 J,$ where $J$ is a constant. We tested five different temperatures, $10^{-1}, 10^{-2}, 10^{-3}, 10^{-4},$ and $10^{-5}$ in units of $\beta J$. We carried out 1,000 independent MC simulations for each temperature, and in each simulation we performed 1.1 million MC steps and neglected the first 0.1 million steps to achieve the system equilibrium (see Fig. \ref{fig:traj} for a representative trajectory).


Fig. \ref{fig:alloystr}c shows the MC free energy difference between the two systems with different defect types as a function of $\beta J$. In the high-temperature regime, the constant term and one-link term (proportional to the number of links) in equation \ref{eq:FE} dominate so that the difference between the two systems is negligible. However, as inverse temperature $\beta J$ increases, the free energy difference between the two systems increases as well. The point defect system has lower free energy as expected, which is consistent with the experimentally known fact that point vacancies have thermal equilibrium concentrations whereas higher-dimensional defects cannot be formed spontaneously \cite{Priester2013}. The entropic effect of multiple vacancy configurations and the stabilizing effect of structure relaxation have previously been used to explain this difference \cite{PhysRevB.67.174105}, but both factors were constant in our simulations so they cannot account for the free energy differences observed here. Also, the numbers of broken bonds are equal, meaning that the ``surface areas'' are the same. Thus, this result implies that the stability of point defects (compared to line and planar defects) is partially due to the lattice topology itself.


\begin{figure}[t]
\subfloat[] {
\includegraphics[scale=0.5]{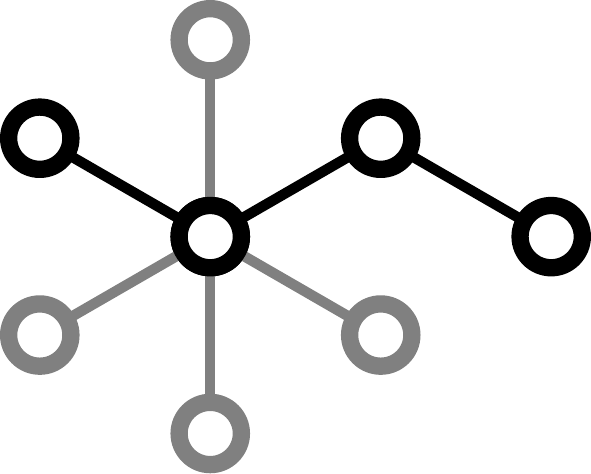}
} \qquad
\subfloat[] {
\includegraphics[scale=0.5]{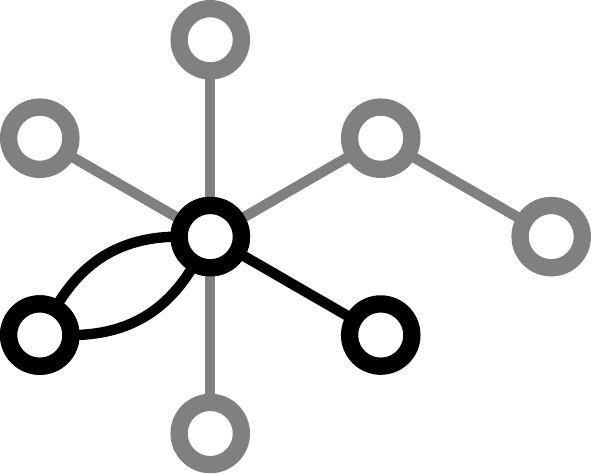}
}
\caption{\label{fig:linex} Examples of subgraphs (black) contributing to the term $[g]$ where $g$ is a 3-link path graph (Fig. \ref{fig:obeta3}a) on the given graph architecture (gray). (b) shows a child graph of a 3-link path graph (Fig. \ref{fig:obeta3}c).}
\end{figure}

We have hitherto considered qualitative differences between different graphs. Can we make a quantitative prediction of the system free energy? Although it is impossible to obtain a general closed form of equation \ref{eq:FE}, a closed form can be obtained for some special types of contributing graphs. We will focus on a path subgraph \footnote{A path graph is a graph where two (terminal) nodes have vertex degree 1 and the other nodes have degree 2.} (see Fig. \ref{fig:linex}a). Among tractable subgraphs, this type of graph has a significant contribution, because it does not have any connected parent graph and $[g] \geq [h]$ if $g$ is a parent graph of $h$, so that the path graph provides one of the largest $[g]$ values among the connected graphs with the same number of links.

For a path subgraph $g$ of length $n(g)$,
\begin{equation}
[g] = \sum A_{i_0 i_1} A_{i_1 i_2} \cdots A_{i_{n(g)-1} i_{n(g)}} = \text{su } A^{n(g)},
\end{equation}
where $\text{su } A$ is defined as the element sum of $A$, \emph{i.e.} $\sum_{ij} A_{ij}$. Note that this term contains contributions from child subgraphs of $g$ (see Fig. \ref{fig:linex}b). Similarly, we can express $E(g)$ for a path subgraph $g$ by a relatively simple form,
\begin{equation}
E(g) = M^{-n(g)-1} \text{su } E^{n(g)},
\end{equation}
and we will define $\tilde{F}_\text{path}(\beta)$ as the summation of free energy contributors corresponding to path graphs:
\begin{equation}
\tilde{F}_\text{path}(\beta) = \sum_{\text{path } g} -\frac{1}{\beta} \frac{(-\beta/2)^{n(g)}}{n(g)!} K(g) E(g) [g].
\end{equation}
By using matrix diagonalization (see \hyperref[subsec:appC]{\textbf{Appendix C}}), it can be shown that
\begin{equation} \label{eq:Fpath}
\tilde{F}_\text{path} (\beta) = \frac{1}{2M^2} \sum_{i,j}^{N,M} \frac{|c_i|^2 |d_j|^2 \lambda_i \mu_j}{1 + \beta \lambda_i \mu_j / M},
\end{equation}
for $Mk_B T > \max(|\lambda_i|) \max(|\mu_j|)$. Here $\{ \lambda_i \}$ and $\{ \mu_j \}$ respectively represent the spectra of $A$ and $\epsilon = E - \sum_{ij} E_{ij}/M^2$, and their corresponding eigenvector sets are respectively $\{ | i \rangle_A \}$ and $\{ | j \rangle_\epsilon \}$. We use inner products $c_i = \langle \mathbf{1} | i \rangle_A$ and $d_j = \langle \mathbf{1} | j \rangle_\epsilon$, by denoting an all-ones vector by $| \mathbf{1} \rangle$.

The free energy formula with the path graph factor is
\begin{widetext}
\begin{equation} \label{eq:final}
F(\beta) = - N k_B T \ln M +\frac{1}{2} E_0 \cdot \text{su } A - \frac{\beta}{4M^2} \left(\text{tr } \epsilon^2 - \frac{2}{M} \text{su } \epsilon^2\right) \text{tr } A^2 + \frac{1}{2M^2} \sum_{i,j} \frac{|c_i|^2 |d_j|^2 \lambda_i \mu_j}{1 + \beta \lambda_i \mu_j / M} + \mathcal{O}(\beta^2),
\end{equation}
where $\text{tr } A$ indicates the trace of $A$, while a simple linear approximation of equation \ref{eq:FE} (see \hyperref[subsec:appD]{\textbf{Appendix D}}) gives
\begin{equation} \label{eq:linear}
F(\beta) = -Nk_B T \ln M + \frac{1}{2} E_0 \cdot \text{su } A - \frac{\beta}{4M^2} \left\{ \left(\text{tr } \epsilon^2 - \frac{2}{M} \text{su } \epsilon^2 \right) \text{tr } A^2 + \frac{2}{M} \text{su } \epsilon^2 \text{su } A^2\right\} + \mathcal{O}(\beta^2).
\end{equation}
\end{widetext}

To illustrate the utility of those approximate formulae, let us consider an example from biophysics. The designability of a protein structure is defined as the number of sequences that fold into the given structure as their lowest energy state. Biophysicists have been used this concept to investigate the principles of protein design and protein evolution (for review, see \cite{desgrev}). As previously discussed \cite{PhysRevLett.90.218101}, there is a strong relationship between designability and sequence space free energy, \emph{i.e.} the free energy of a heteropolymer in sequence space, instead of conformation space.


The Hamiltonian of a protein structure is given by
\begin{equation}
\mathcal{H} = \frac{1}{2} \sum_{i,j}^{N,N} A_{ij} E_{\text{AA}(i) \text{AA}(j)}.
\end{equation}
Here $A$ is called a contact matrix in the protein structure literature. Each element of $A$, $A_{ij}$, is 1 when residues $i$ and $j$ are nearest neighbors on the lattice but not adjacent on the protein backbone, and $A_{ij} = 0$ otherwise. $E$ is an interaction matrix that contains interaction energies for every pair of amino acid types. $N$ is the chain length, and $\text{AA}(k)$ is the amino acid type of residue $k$. This formula is analogous to the Hamiltonain for a spin model (equation \ref{eq:hamil}; see \cite{Shakhnovich01111993}), so we can apply equation \ref{eq:final}, or equation \ref{eq:linear} to calculate the sequence space free energy at high sequence space temperature (where mutations can occur relatively frequently).

\begin{figure}[tb]
\includegraphics[width=0.5\textwidth]{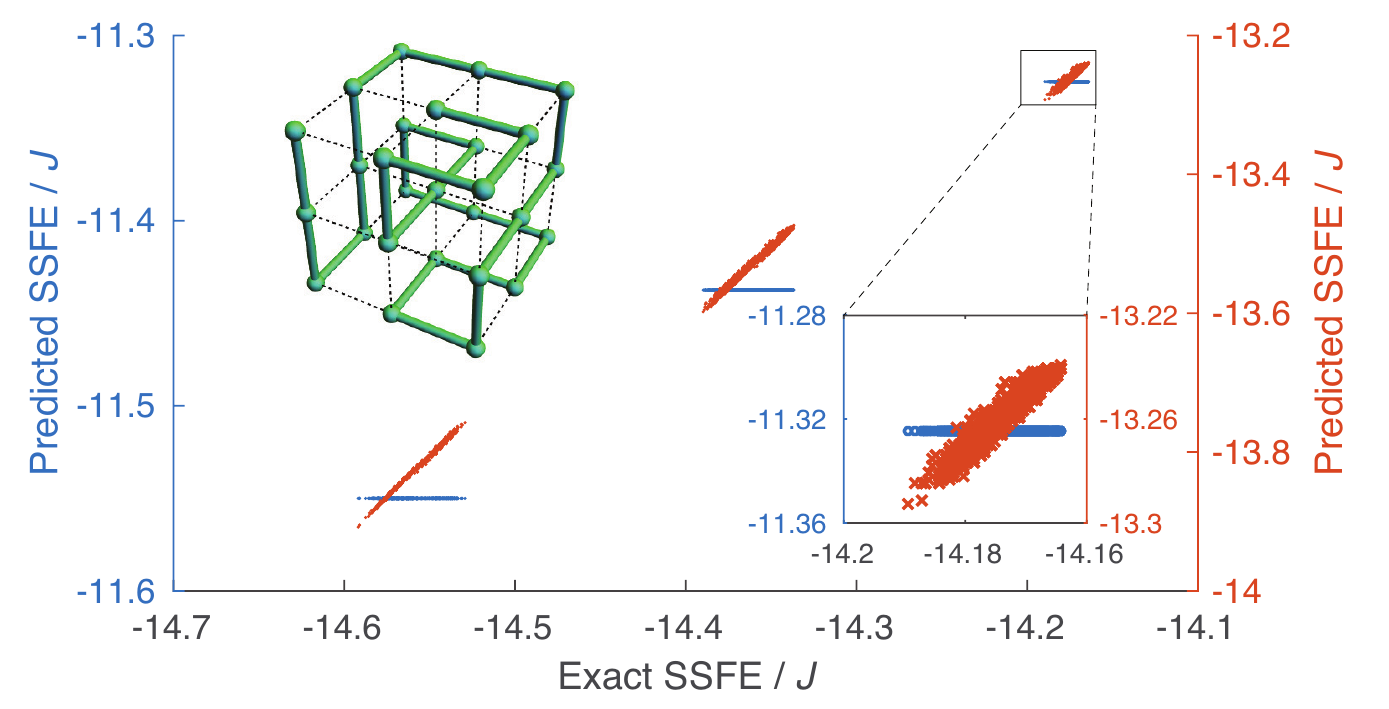}
\caption{\label{fig:scatt} Scatter plots of sequence space free energy (SSFE) distributions for 10,000 lattice protein structures, comparing exact and predicted SSFEs at $\beta J = 0.1$. Equation \ref{eq:final} (red) and equation \ref{eq:linear} (blue) were used to calculate predicted SSFEs. (Upper left) a cartoon of a prototypical lattice protein. (Lower right) zoom of the boxed region.}
\end{figure}

We studied the sequence space of a 3$\times$3$\times$3 lattice protein structure (Fig. \ref{fig:scatt}, upper left), whose intra-chain interaction network can be represented by a graph with 27 nodes and 28 links (the total number of non-covalent contacts). We used 10,000 representative structures among 103,346 maximally compact structures \cite{JCP1990} to reduce the computational cost, following Heo \emph{et al} \cite{Heo:2011ys}. We also employed two types of amino acids, and the interaction matrix represents hydrophobic and polar interactions:
\begin{equation}
E = \left[\begin{array}{cc}
-3J & J \\
J & 0\end{array}\right]
\end{equation}
We designated an arbitrary chosen conformation as ``native structure'' and scanned all $2^{27}$ (about 1.3$\times 10^8$) possible sequences to compute $Z(\beta) = \sum_{\text{sequences}} e^{-\beta \mathcal{H}}$ and the corresponding $F(\beta)$ for a given $\beta$. We denote this $F$ as the ``exact sequence space free energy,'' to distinguish it from a prediction made using equation \ref{eq:final} or \ref{eq:linear}, which we will call the ``predicted sequence space free energy.'' We repeated this calculation for all 10,000 structures.

Fig. \ref{fig:scatt} shows the sequence space free energy distributions over 10,000 lattice proteins at temperature $\beta J = 0.1$. The predicted values from equation \ref{eq:final} correlate strongly with the exact values (red, right axis), while predictions using the simpler approximation, equation \ref{eq:linear}, are not capable of discriminating between structures with different sequence space free energies (blue, left axis). The deficiency of equation \ref{eq:linear} is due to degeneracies in $\text{su } A^2$ and $\text{tr } A^2$, which equation \ref{eq:linear} cannot resolve. However, even in the former case, strict one-to-one correspondence does not hold between the exact and predicted values (lower right), because of contributions from higher-order terms that $\tilde{F}_\text{path}(\beta)$ does not capture. Note that the structures are mainly grouped by three different $\text{su } A^2$ values, implying that the system is still in the high-temperature regime, where higher-order terms do not dominate.


In this Letter, we presented an analytical method for calculating the free energy of a spin model on a simple graph. Through this approach, we find that the free energy contribution of the graph topology, realized by products of adjacency matrix elements, can be separated from energetic factors. The theory was illustrated by comparing chain and star graphs. Without specifying the interaction matrix, we showed that the star graphs are more stable than chain graphs in the high-temperature regime. The approach was then applied to lattice models of alloys with different defect types, which lead to different free energies, even when the systems had the same defect surface areas. We also showed that linear graphs are special in the sense that their infinite sum can be computed exactly, and this approach was applied to the protein design problem. The relative order of sequence space free energies of lattice proteins was predicted relatively accurately by the formula containing the infinite sum from the linear graph contribution, whereas a mere linear approximation could not discriminate among structures with the same $\text{su } A^2$ values. We hope that this theory will be expanded and applied to other graph-related problems in physics, from more complex spin systems to biological systems and also social networks.

\begin{acknowledgments}
We appreciate valuable comments from Erel Levine and Kunok Chang. This work is supported by NIH grant GM068670.
\end{acknowledgments}

\bibliography{FreeEnergy-draft}{}

\clearpage

\setcounter{page}{1}
\setcounter{figure}{0}
\setcounter{equation}{0}
\renewcommand{\thepage}{S\arabic{page}}
\renewcommand{\theequation}{S\arabic{equation}}
\renewcommand{\thefigure}{S\arabic{figure}}

\section{Supplemental Material}
\subsection{Appendix A: Derivation of Equations \ref{eq:Zorig} and \ref{eq:Hg}}
\label{subsec:appA}
As described in the text, the Hamiltonian and partition function are respectively given by
\begin{eqnarray}
\mathcal{H} &=& \frac{1}{2} \sum_{i,j}^{N,N} A_{ij} E_{s(i) s(j)} \\ \label{eq:PForig}
Z(\beta) &=& \sum_{\{s\}} 1 - \beta \sum_{\{s\}} \mathcal{H} + \frac{\beta^2}{2!} \sum_{\{s\}} \mathcal{H}^2 - \cdots.
\end{eqnarray}

The first summation in $Z(\beta)$ is simply the number of all possible state configurations: $M^N$. This is a purely entropic term. Moving to the $\mathcal{O}(\beta)$ term, each link (i.e. nonzero $A_{ij}$) contributes an energetic contribution of $\sum_{s(i), s(j)} E_{s(i) s(j)}$, while the remaining nodes (other than $i$ and $j$) contribute entropically by $M^{N-2}$. In other words,
\begin{eqnarray}
\sum_{\{s\}} \mathcal{H} &=& \frac{1}{2} M^{N-2} \sum_{i,j} A_{ij} \sum_{k,l} E_{kl} \\ \label{eq:mean-field2}
&=& \frac{1}{2} M^N \sum_{i,j} A_{ij} E_0,
\end{eqnarray}
where $E_0$ is the average energy. Noting that $\sum A_{ij}/2$ is the total number of links, equation \ref{eq:mean-field2} describes the mean-field energetic contribution.

\begin{figure}[tb]
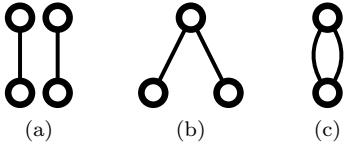

\subfloat[] {
\includegraphics[scale=0.5]{Fig_Obeta2a}
} \qquad
\subfloat[] {
\includegraphics[scale=0.5]{Fig_Obeta2b}
} \qquad
\subfloat[] {
\includegraphics[scale=0.5]{Fig_Obeta2c}
}
\caption{\label{fig:obeta2} Diagrammatical representations for three different types of two-link multigraphs.}
\end{figure}

Next, let us explicitly calculate the $\mathcal{O}(\beta^2)$ term in equation \ref{eq:PForig}:
\begin{equation} \label{eq:beta2}
\frac{\beta^2}{2!} \sum_{\{s\}} \mathcal{H}^2 = \frac{\beta^2}{2! \cdot 4} \sum_{ijkl} A_{ij} A_{kl} \sum_{\{s\}} E_{s(i) s(j)} E_{s(k) s(l)},
\end{equation}
and there are three different types of energetic contributions, depending on the relationship between the two node pairs $(i,j)$ and $(k,l)$. The contribution of each type is represented diagrammatically in Fig. \ref{fig:obeta2}, where each link represents a single energy term. Note that the graphs are no more simple in these diagrams; they are \emph{multigraphs}, which allow multiple links.

To be particular, Fig. \ref{fig:obeta2} describes three different cases: (a) the pairs are totally disconnected (no nodes are same), in which case energetic contribution is $M^{N-4} \sum_{m,n,p,q} E_{mn} E_{pq}$. (b) They share only one of their nodes ($i=k$ or $i=l$ or $j=k$ or $j=l$; other nodes are all different). The contribution is $M^{N-3} \sum_{m,n,p} E_{mn} E_{np}$. (c) The two pairs are identical (either $i=k$ and $j=l$, or $i=l$ and $j=k$). The contribution to the partition function is $M^{N-2} \sum_{m,n} E_{mn}^2$. This is where our general definition of $E(g)$ in the main text came from:
\begin{equation}
E(g) = M^{-n(\text{nodes})} \sum_\text{nodes} \prod_{k=1}^{n(g)} E_{l_k},
\end{equation}
which is the energetic contribution of each multigraph normalized by the entropic contribution $M^N$.

The number of possible $(i,j,k,l)$ combinations for each multigraph type should be calculated. Let us use the definition of $[g]$ in the main text:
\begin{equation}
[g] = \sum_\text{nodes} \prod_{k=1}^{n(g)} A_{l_k},
\end{equation}
and the three different two-link multigraphs (Fig. \ref{fig:obeta2}) give
\begin{eqnarray} \label{eq:bra2a}
\left[\graphtwoa\right] &=& \sum_{i,j,k,l} A_{ij} A_{kl} \\ \label{eq:bra2b}
\left[\graphtwob\right] &=& \sum_{i,j,k} A_{ij} A_{jk} \\ \label{eq:bra2c}
\left[\graphtwoc\right] &=& \sum_{i,j} A_{ij}^2
\end{eqnarray}
However, they do not count the exact numbers, because equation \ref{eq:bra2a} contains contributions from equations \ref{eq:bra2b} and \ref{eq:bra2c}, and equation \ref{eq:bra2b} includes those from equation \ref{eq:bra2c}. The exact number of contributing combinations for graph type $g$, which will be denoted as $W(g)$, is given as follows:
\begin{eqnarray} \label{eq:W2c}
W\left(\graphtwoc\right) &=& 2 \left[\graphtwoc\right] \\ \label{eq:W2b}
W\left(\graphtwob\right) &=& 4\left\{ \left[\graphtwob\right] - W\left(\graphtwoc\right) \right\} \\ \label{eq:W2a}
W\left(\graphtwoa\right) &=& \left[\graphtwoa\right] - W\left(\graphtwob\right)  - W\left(\graphtwoc\right)
\end{eqnarray}
The factors of 2 and 4 respectively in equations \ref{eq:W2c} and \ref{eq:W2b} come from the symmetry counting.

Furthermore, the energy contribution can be decomposed by defining energy deviation $\epsilon_{kl} = E_{kl} - E_0$. Considering that $\sum_{k,l} \epsilon_{kl} = 0$, arithmetics leads to
\begin{eqnarray}
\sum_{k,l} E_{kl}^2 &=& M^2 E_0^2 + \sum_{k,l} \epsilon_{kl}^2, \\
\sum_{k,l,m} E_{kl} E_{lm} &=& M^3 E_0^2 + \sum_{k,l,m} \epsilon_{kl} \epsilon_{lm}, \\
\sum_{k,l,m,n} E_{kl} E_{mn} &=& M^4 E_0^2.
\end{eqnarray}

\begin{widetext}
Therefore, the explicit form of the $\mathcal{O}(\beta^2)$ term (equation \ref{eq:beta2}) is
\begin{equation}
\frac{\beta^2}{2! \cdot 4} \left[ W\left(\graphtwoa\right) M^{N-4} \cdot M^4 E_0^2 + W\left(\graphtwob\right) M^{N-3} \left(M^3 E_0^2 + \sum_{k,l,m} \epsilon_{kl} \epsilon_{lm} \right) + W\left(\graphtwoc\right) M^{N-2} \left(M^2 E_0^2 + \sum_{k,l} \epsilon_{kl}^2 \right) \right],
\end{equation}
which becomes
\begin{equation}
\frac{\beta^2}{2! \cdot 4} M^N \left\{ E_0^2 \left[ W\left(\graphtwoa\right) + W\left(\graphtwob\right) + W\left(\graphtwoc\right) \right] + W\left(\graphtwob\right) M^{-3} \sum_{k,l,m} \epsilon_{kl} \epsilon_{lm} + W\left(\graphtwoc\right) M^{-2} \sum_{k,l} \epsilon_{kl}^2 \right\},
\end{equation}
or, using equations \ref{eq:W2c}, \ref{eq:W2b}, and \ref{eq:W2a},
\begin{equation}
\frac{\beta^2}{2! \cdot 4} M^N \left\{ E_0^2 \left[ \graphtwoa \right] + 4M^{-3} \sum_{k,l,m} \epsilon_{kl} \epsilon_{lm} \left[ \graphtwob \right] + \left(2M^{-2} \sum_{k,l} \epsilon_{kl}^2 -8M^{-3} \sum_{k,l,m} \epsilon_{kl} \epsilon_{lm} \right) \left[ \graphtwoc \right] \right\}.
\end{equation}
Note that the first term in parentheses is indeed
\begin{equation}
E_0^2 \left[\graphtwoa\right] = E_0^2 (\sum A_{ij})^2 = E_0^2 \left[\graphone\right]^2 = \left\{E_0 W\left(\graphone\right)\right\}^2.
\end{equation}
\end{widetext}

To systematically expand this approach to higher-order terms, we will use the node contraction operation. As shown above, the critical step is calculation of $W(g)$, the exact degeneracy of graph $g$. Once we know $W(g)$, the partition function can be written in the form of
\begin{equation} \label{eq:pf}
Z(\beta) = M^N \left\{1 + \sum_g \frac{(-\beta/2)^{n(g)}}{n(g)!} W(g) E(g)\right\},
\end{equation}
where $n(g)$ is the number of links in graph $g$ and the summation is over all possible (connected and disconnected) graphs $g$. As defined in the main text, let us consider child and parent graphs: a child graph $h$ of graph $g$ is obtained by contraction of unconnected nodes in graph $g$, and $g$ is called a parent graph of $h$. Note that a child graph has the same number of links as its parent. Then, generally the following equation holds:
\begin{equation} \label{eq:Wg}
W(g) = K(g) \left\{[g] + \sum_{g' \in \mathcal{C}(g)} (-1)^{m(g',g)} K(g',g) [g']\right\},
\end{equation}
where $K(g)$ is the combinatoric factor to construct graph $g$ from $n(g)$ links, $K(g', g)$ is the combinatoric factor to generate graph $g'$ from graph $g$ by node contraction, $m(g', g)$ is the minimal number of contraction operations required to construct $g'$ from $g$, and $\mathcal{C}(g)$ is the set containing all child graphs of graph $g$. Since the number of links is same for parent and child graphs, we can write equation \ref{eq:pf} in terms of $[g]$ by using the following formula: for the given number of links $n$,
\begin{equation} \label{eq:WEHg}
\sum_{\substack{g \text{ \emph{s.t.}} \\ n(g) = n}} W(g) E(g) = \sum_{\substack{g \text{ \emph{s.t.}} \\ n(g) = n}} H(g) [g],
\end{equation}
where
\begin{eqnarray} \nonumber
H(g) &=& K(g) E(g) \\ \label{eq:Hg2}
&& +\sum_{g' \in \mathcal{P}(g)} (-1)^{m(g,g')} K(g,g') K(g') E(g'),
\end{eqnarray}
and $\mathcal{P}(g)$ is a set containing all parents of graph $g$. This is equation \ref{eq:Hg} in the main text, and we can convert equation \ref{eq:pf} into equation \ref{eq:Zorig} in the main text:
\begin{equation}
Z(\beta) = M^N \left\{1 + \sum_g \frac{(-\beta/2)^{n(g)}}{n(g)!} H(g) [g]\right\},
\end{equation}

Note that since
\begin{equation}
K(g) + \sum_{g'} (-1)^{m(g,g')} K(g,g') K(g') = 0
\end{equation}
unless $g$ is a one-node graph, we can always replace $E(g)$ in equation \ref{eq:Hg2} by $\epsilon(g)$, which is defined as
\begin{equation}
\epsilon(g) = M^{-n(\text{nodes})} \sum_\text{nodes} \prod_{k=1}^{n(g)} (E_{l_k} - E_0),
\end{equation}
which will be used in \hyperref[subsec:appC]{\textbf{Appendix C}}.

\subsection{Appendix B: The Chain and Star Graph Systems}
\label{subsec:appB}

\begin{figure}[tb]
\subfloat[] {
\includegraphics[scale=0.5]{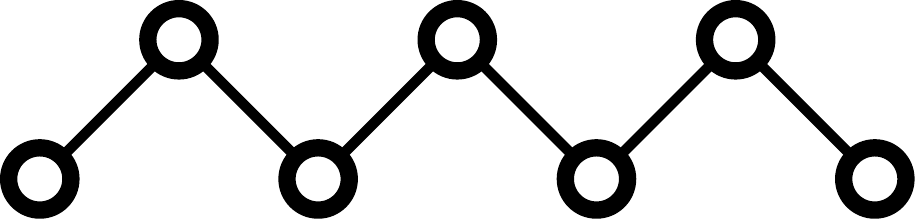}
} \qquad
\subfloat[] {
\includegraphics[scale=0.5]{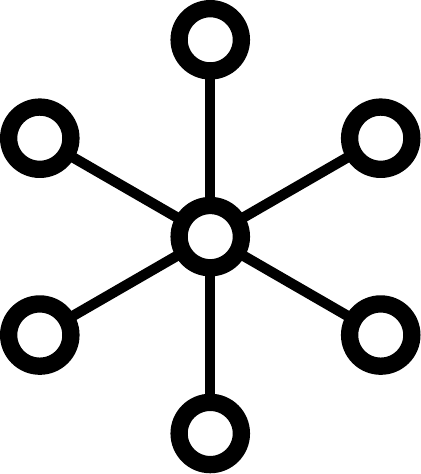}
}
\caption{\label{fig:chainstar} The 6-link chain and star graphs.}
\end{figure}

In the main text, we introduced the chain and star graphs (Fig. \ref{fig:chainstar}). We will show that $[g]^\text{chain} < [g]^\text{star}$ in general, and also investigate the Ising models on both graphs.

The elements of their adjacency matrices for the chain and star graph systems, denoted as $A^\text{chain}$ and $A^\text{star}$ respectively, are given below:
\begin{eqnarray}
A_{ij}^\text{chain} &=& \delta_{i,j+1} + \delta_{i,j-1} \\
A_{ij}^\text{star} &=& \delta_{i,1} + \delta_{j,1} - 2 \delta_{i,1} \delta_{j,1},
\end{eqnarray}
where node index 1 for the star graph indicates the center node. In calculating $[g]$ for a multiple graph $g$, we have a summation over various products of $A_{ij}$, and when there is a common index (\emph{e.g.} $j$ in $A_{ij} A_{jk}$), the star graph gives a larger contribution to $[g]$ than the chain graph, since the two indices are disentangled in the former system. For example, let us calculate $S = \sum_{ijk} A_{ij} A_{jk}$ for both cases:
\begin{eqnarray} \nonumber
S^\text{chain} &=& \sum_{ijk}^N (\delta_{i-1,j} + \delta_{i+1,j})(\delta_{j,k+1} + \delta_{j,k-1}) \\ \nonumber
&=& \sum_{i,k} (\delta_{i-1,k+1} + \delta_{i-1,k-1} + \delta_{i+1,k+1} + \delta_{i+1,k-1}) \\ \nonumber
&=& (N-2) + (N-1) + (N-1) + (N-2) \\
&=& 4N-6,
\end{eqnarray}
where $k$ values in $(N-k)$ terms are obtained by considering the boundary conditions. Also,
\begin{eqnarray} \nonumber
S^\text{star} &=& \sum_{ijk}^N (\delta_{i,1} + \delta_{j,1}  - 2 \delta_{i,1} \delta_{j,1})(\delta_{j,1} + \delta_{k,1}  - 2 \delta_{j,1} \delta_{k,1}) \\ \nonumber
&=& \sum_{ijk} (\delta_{i,1} \delta_{j,1} + \delta_{i,1} \delta_{k,1} + \delta_{j,1}^2 + \delta_{j,1} \delta_{k,1} + \mathcal{O}(\delta^3)) \\ \nonumber
&=& N^2 + 3N - 4N \\
&=& N^2 - N.
\end{eqnarray}
Here, since $\delta_{ij}^2 = \delta_{ij}$, we get a quadratic dependence on $N$, which does not appear in the first case. Hence, $S^\text{chain} < S^\text{star}$ for $N > 3$. We can use a similar argument to show that
\begin{equation}
[g]^\text{chain} < [g]^\text{star},
\end{equation}
in general.

The Ising model can be applied to these two graph systems. Let the energy matrix be
\begin{equation}
\beta E = \left( \begin{array}{cc}
-J & J \\
J & -J
\end{array}\right),
\end{equation}
where $J > 0$. The partition function for the Ising chain (with open boundary conditions) is well known: for the chain length $N \geq 2$,
\begin{equation}
Z^\text{chain} = 2(2\cosh J)^{N-1}.
\end{equation}
For the ``Ising star,'' calculation of the partition function is straightforward:
\begin{eqnarray}
Z^\text{star} &=& \sum_{\{s_i\}} e^{J \sum_{\langle i, j \rangle} s_i s_j} \\
&=& \sum_{\{s_i\}} e^{J \sum_{i \neq 1} s_1 s_i} \\
&=& \sum_{s_1 = \pm 1} \prod_{i \neq 1} \sum_{s_i = \pm 1} e^{J s_1 s_i} \\
&=& (4 \cosh J)^{N-1}.
\end{eqnarray}

Hence, for $N=2$, the two systems have the same free energy (and same graph structure), but for $N > 2$, $Z^\text{chain} < Z^\text{star}$ so that the Ising star is more stable than the Ising chain at any temperature.

\subsection{Appendix C: Derivation of Equation \ref{eq:Fpath}}
\label{subsec:appC}

In the main text, we define
\begin{equation}
\tilde{F}_\text{path}(\beta) = \sum_{\text{linear } g} -\frac{1}{\beta} \frac{(-\beta/2)^{n(g)}}{n(g)!} K(g) E(g) [g],
\end{equation}
and show
\begin{eqnarray}
[g] &=& \text{su } A^{n(g)}, \\
E(g) &=& M^{-n(g)-1} \text{su } E^{n(g)},
\end{eqnarray}
where $\text{su } A \equiv \sum_{ij} A_{ij}$. Also,
\begin{equation}
K(g) = 2^{n(g)-1} \cdot n(g)!,
\end{equation}
where $2^n \cdot n!$ is a combinatoric factor and due to symmetry we have the double-counting correction of 1/2. Therefore, substitution leads to
\begin{equation}
\tilde{F}_\text{path}(\beta) = -\frac{1}{2\beta M} \sum_{\substack{\text{path } g \\ \text{\emph{s.t.} } n(g) \geq 2}} \left(-\frac{\beta}{M}\right)^{n(g)} \text{su } E^{n(g)} \text{su } A^{n(g)}.
\end{equation}
As shown at the end of \hyperref[subsec:appA]{\textbf{Appendix A}}, we can rewrite the previous equation by
\begin{equation} \label{eq:Flinprim}
\tilde{F}_\text{path}(\beta) = -\frac{1}{2\beta M} \sum_{\substack{\text{path } g \\ \text{\emph{s.t.} } n(g) \geq 2}} \left(-\frac{\beta}{M}\right)^{n(g)} \text{su } \epsilon^{n(g)} \text{su } A^{n(g)},
\end{equation}
where $\epsilon = E - E_0 I$.

Diagonalization helps to get a closed form of equation \ref{eq:Flinprim}. For eigenvalues $\lambda_1 \ge \lambda_2 \ge \cdots \ge \lambda_N$ of square matrix $B$ of size $N$, we have $\text{su } B^k = \sum_{i=1}^N |a_i|^2 \lambda_i^k,$ where $a_i$ is the inner product of the eigenvector corresponding to eigenvalue $\lambda_i$ and an all-ones vector of size $N$. Using this, we can write
\begin{eqnarray}
\text{su } A^k &=& \sum_{i=1}^N |c_i|^2 \lambda_i^k \\
\text{su } \epsilon^k &=& \sum_{j=1}^M |d_j|^2 \mu_j^k.
\end{eqnarray}
where $\{ \lambda_i \}$ and $\{ \mu_i \}$ represent the spectra of $A$ and $\epsilon$ respectively, and $\{ c_i \}$ and $\{ d_i \}$ correspond to $\{ a_i \}$ above for $A$ and $\epsilon$ respectively. Thus, equation \ref{eq:Flinprim} becomes
\begin{equation}
\tilde{F}_\text{path} (\beta) = \frac{1}{2M^2} \sum_{i,j}^{N,M} \frac{|c_i|^2 |d_j|^2 \lambda_i \mu_j}{1 + \beta \lambda_i \mu_j / M},
\end{equation}
for $Mk_B T > \max(|\lambda_i|) \max(|\mu_j|)$.

\subsection{Appendix D: Derivation of Equation \ref{eq:linear}}
\label{subsec:appD}

As discussed in the main text, the free energy is given by
\begin{equation}
F(\beta) = -Nk_B T \ln M + \sum_{\text{connected } g} \tilde{F}(g,\beta),
\end{equation}
where
\begin{equation}
\tilde{F}(g, \beta) = -\frac{1}{\beta} \frac{(-\beta/2)^{n(g)}}{n(g)!} H(g) [g].
\end{equation}

Equation \ref{eq:linear} is obtained by termination of the series at the linear order $\mathcal{O}(\beta)$, \emph{i.e.}, considering only the terms with $n(g) \leq 2$:
\begin{equation} \label{eq:Flin}
F_\text{approx}(\beta) = -Nk_B T \ln M + \sum_{\substack{\text{connected } g \\ \text{\emph{s.t.} } n(g) \leq 2}} \tilde{F}(g,\beta),
\end{equation}
where, using equation \ref{eq:WEHg},
\begin{equation}
\tilde{F}(g, \beta) = -\frac{1}{\beta} \frac{(-\beta/2)^{n(g)}}{n(g)!} W(g) E(g).
\end{equation}
For a one-link graph, we have
\begin{equation}
W\left(\graphone\right) = \left[\graphone\right] = \sum_{ij} A_{ij},
\end{equation}
and
\begin{equation}
E\left(\graphone\right) = H\left(\graphone\right) = \sum_{kl} E_{kl} = E_0.
\end{equation}
Plugging these and equations \ref{eq:W2c} and \ref{eq:W2b} into equation \ref{eq:Flin}, we get equation \ref{eq:linear} in the main text.\newline\newline

\subsection{Appendix E: Monte Carlo Trajectories}
\label{subsec:appE}

\begin{figure}[b]
\includegraphics[width=0.5\textwidth]{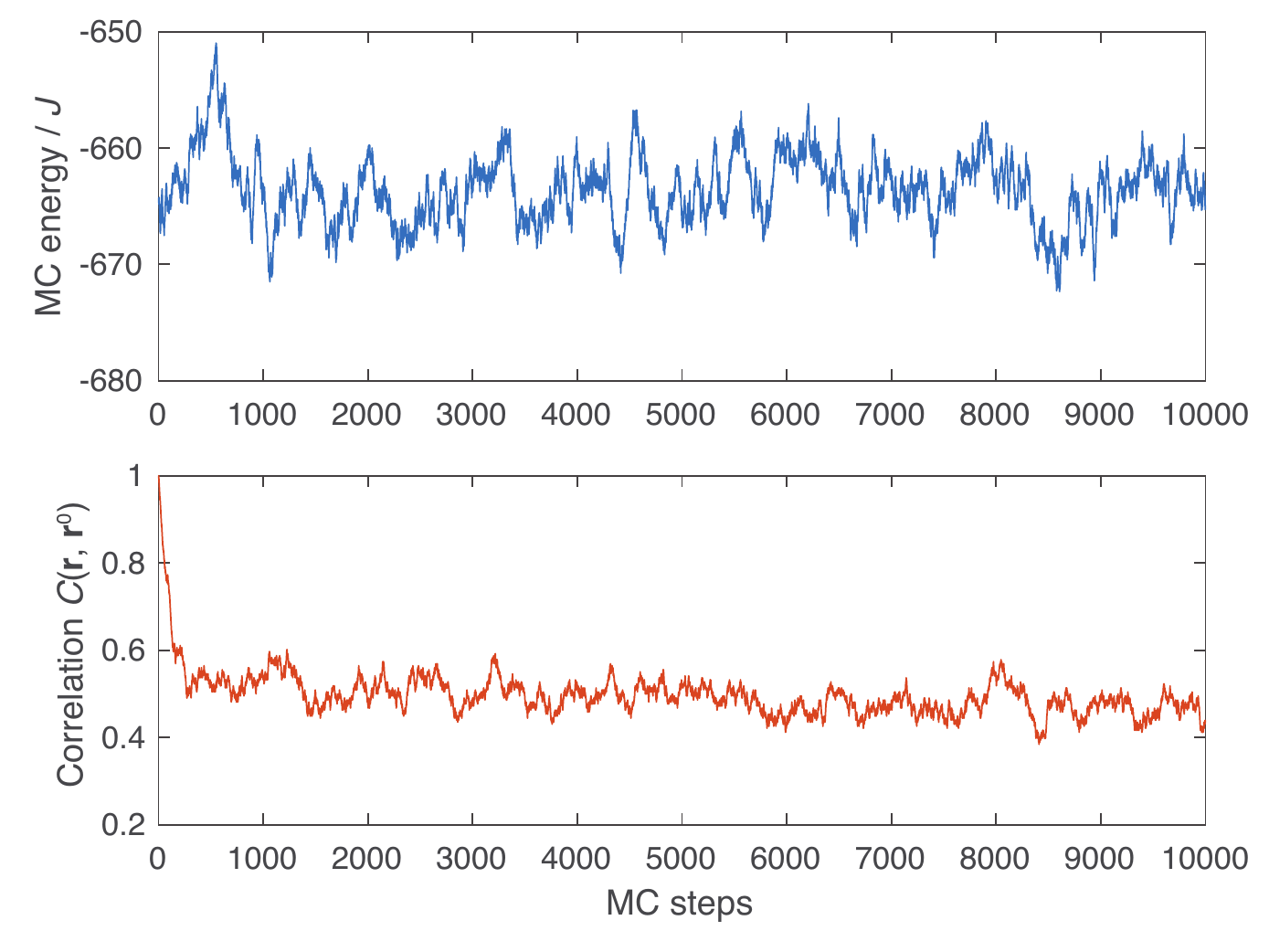}
\caption{\label{fig:traj} Trajectory plots of MC energy (top) and a correlation function (bottom) to the initial configuration, for a system with $\beta J = 0.1$.}
\end{figure}

Figure \ref{fig:traj} shows a set of trajectories of MC energy and a correlation function to the initial configuration, for a system with $\beta J = 0.1$. As the correlation function trajectory suggests, the system gets equilibrated relatively quickly, implying that the system temperature is comparably high. This is consistent with the assumption throughout this Letter.

\end{document}